\documentclass[twoside,reqno,12pt]{amsart}
\usepackage{amssymb}
\usepackage{enumerate}
\usepackage{a4wide,natbib}
\usepackage[english]{babel}
\usepackage{amsmath}
\usepackage{graphicx,algorithm,algorithmic,url}
\begin{document}
\begin{flushleft}
{\Large \bf GSGS: A Computational Framework to Reconstruct Signaling Pathways from Gene Sets}
\vskip2em
Lipi Acharya$^{\S}$, Thair Judeh$^{*,\dag}$, Zhansheng Duan$^{\flat}$, Michael Rabbat$^{\ddag}$ and Dongxiao Zhu$^{*,\dag,\P}$
\vskip1em
$^{*}$Department of Computer Science, Wayne State University, 5057 Woodward Avenue, Detroit, MI 48202, USA.
\vskip1em
$^{\S}$Department of Computer Science, University of New Orleans, 2000 Lakeshore Drive, New Orleans, LA 70148, USA.
\vskip1em
$^{\dag}$Research Institute for Children, Children's Hospital, New Orleans, LA 70118, USA.
\vskip1em
$^{\flat}$Center for Information Engineering Science Research, Xi'an Jiaotong University, 28 Xianning West Road, Xi'an, Shaanxi 710049, China.
\vskip1em
$^{\ddag}$Department of Electrical and Computer Engineering, McGill University, 3480 University Street, Montr\'{e}al, Qu\'{e}bec H3A 2A7,               Canada.
\vskip1em
$^{\P}$To whom correspondence should be addressed.
\footnotetext[1]{A major portion of this submission is IEEE-copyrighted.}
\end{flushleft}

\section*{abstract}
We propose a novel two-stage Gene Set Gibbs Sampling (GSGS) framework, to reverse engineer signaling pathways from gene sets inferred from molecular profiling data. We hypothesize that signaling pathways are structurally an ensemble of overlapping linear signal transduction events which we encode as Information Flow Gene Sets (IFGS's). We infer pathways from gene sets corresponding to these events subjected to a random permutation of genes within each set. In Stage I, we use a source separation algorithm to derive unordered and overlapping IFGS's from molecular profiling data, allowing cross talk among IFGS's. In Stage II, we develop a Gibbs sampling like algorithm, Gene Set Gibbs Sampler, to reconstruct signaling pathways from the latent IFGS's derived in Stage I. The novelty of this framework lies in the seamless integration of the two stages and the hypothesis of IFGS's as the basic building blocks for signal pathways. In the proof-of-concept studies, our approach is shown to outperform the existing Bayesian network approaches using both continuous and discrete data generated from benchmark networks in the DREAM initiative. We perform a comprehensive sensitivity analysis to assess the robustness of the approach. Finally, we implement the GSGS framework to reconstruct signaling pathways in breast cancer cells.

\section {Introduction}
\label{sec:1}
A central goal of computational systems biology is to decipher signal transduction pathways in living cells. Characterization of complicated interaction patterns in signaling pathways can provide insights into biomolecular interaction and regulation mechanisms. Consequently, there have been a large body of computational efforts addressing the problem of signaling pathway reconstruction by using Probabilistic Boolean Networks (PBNs) (\citealt{Shmulevich02}, \citealt{Shmulevich03}), Bayesian Networks (BNs) (\citealt{Friedman00}, \citealt{Segal03}, \citealt{Song09}), Relevance Networks (RNs) (\citealt{Butte03}), Graphical Gaussian Models (GGMs) (\citealt{Kishino00}, \citealt{Dobra04}, \citealt{Schaffer05}) and other approaches (\citealt{Gardner03}, \citealt{Tenger03}, \citealt{Altay10a}).

\par Although the existing approaches are useful, they often represent a phenomenological graph of the observed data. For example, parent set of each gene in case of BNs, indicates statistically causal relationships. RNs, GGMs and PBNs are computationally tractable even for large signaling pathways, however co-expression criteria used in RNs and GGMs only models a possible functional relevancy, and the use of boolean functions in PBNs may lead to an oversimplification of the underlying gene regulatory mechanisms. Moreover, the aforementioned approaches purely rely on molecular profiling data generated from high-throughput platforms, which are often noisy with high experimental cost associated with them. Consequently, the reconstructed networks may fail to represent the underlying signal transduction mechanisms.

\par On the other hand, gene set based analysis has received much attention in recent years. An initial characterization of large-scale molecular profiling data often results in the identification of many pathway components, which we refer to as gene sets. Availability of several computational and experimental strategies have led to a rapid accumulation of gene sets in the biomedical databases. A gene set compendium is comprised of a large number of overlapping gene sets as each gene may simultaneously participate in many biological processes. Overlapping reflects the interconnectedness among gene sets and should be exploited to infer the underlying gene regulatory network. Our motivation of considering a gene set based approach for network reconstruction falls into many other categories. For instance, a gene set based approach can more naturally incorporate higher order interaction mechanisms as opposed to individual genes. In comparison to molecular profiling data, gene sets are more robust to noise and facilitate data integration from multiple data acquisition platforms. A gene set based approach can allow us to explicitly consider signal transduction mechanisms underlying individual gene sets. Overall, gene sets provide a rich source of data to infer signaling pathways. The relative advantages of working with gene sets in bioinformatics analyses have been adequately demonstrated (\citealt{Subra05}, \citealt{Pang06}, \citealt{Pang08}, \citealt{Richards10}). However, signaling pathway reconstruction by sufficiently exploiting gene sets, a promising area of bioinformatics research, remains underdeveloped.

\par With few exceptions, the existing network reconstruction approaches do not accommodate gene sets. The frequency method in (\citealt{Rabbat05}) assigns an order to a gene set by assuming a tree structure in the paths between pairs of nodes. However, the method is subjected to fail in the presence of multiple paths between the same pair of nodes. To capture the underlying relations between nodes, the cGraph algorithm presented in (\citealt{Kubica03}) adds weighted edges between each pair of nodes that appear in some gene set. The networks inferred by this approach often contain a large number of false positives. It is also difficult to incorporate prior knowledge about regulator-target pairs in the approaches mentioned above. The EM approach in (\citealt{Zhu06}, \citealt{Rabbat08}) treats permutations of genes in a gene set as missing data and assumes a linear arrangement of genes in each set. Nevertheless, it is necessary to develop a systems biology framework integrating both, identification of significant gene sets and signaling pathways reconstruction from gene sets.

\par A central aspect of developing such network reconstruction frameworks is to understand the structure of signaling pathways. Signaling pathways are an ensemble of several overlapping signaling transduction events with a linear arrangement of genes in each event. We denote these events as Information Flows (IF's). Information Flow Gene Sets (IFGS's) stand for the gene sets obtained by randomly permuting the order of genes in each IF. Thus, an IF and an IFGS share the same set of genes, however the latter lacks gene ordering information or it is \emph{unordered}.  We hypothesize that IF's form the building blocks for signaling pathways and uniquely determine their structures. One plausible way to retrieve the latent, unordered and overlapping IFGS's from molecular profiling data is to use source separation approaches, such as Singular Value Decomposition (SVD) (Stage I). The true signaling pathways can be reconstructed by inferring a distribution of more likely orders of the genes in each IFGS (Stage II).

\par In this paper, we design a two-stage Gene Set Gibbs Sampling (GSGS) framework by seamlessly integrating deconvolution of IFGS's and signaling pathway reconstruction from IFGS's. In Stage I, we infer unordered and overlapping IFGS's corresponding to the latent signal transduction events. In Stage II, we develop a stochastic algorithm Gene Set Gibbs Sampler under the Gibbs sampling framework (\citealt{Gelman03}, \citealt{Givens05}) to reconstruct signal pathways from IFGS's inferred in Stage I. The algorithm treats the ordering of genes in an IFGS as a random variable, and samples signaling pathways from the joint distribution of IFGS's. The two-stage GSGS framework is novel from various aspects, such as the hypothesis of IFGS's as the basic building blocks for signal pathways, the definition of gene orderings as a random variable to accommodate higher-order interaction as opposed to individual gene expression, and probabilistic network inferences.
\par We comprehensively examine the performance of our approach by using two gold standard networks from DREAM (Dialogue for Reverse Engineering Assessments and Methods) initiative and compare it with the Bayesian network approaches K2 (\citealt{Cooper92}, \citealt{Murphy01b}b) and MCMC (\citealt{Murphy01a}a, \citealt{Murphy01b}b). We also perform sensitivity analysis to access the robustness of the framework to the under-sampling and over-sampling of gene sets. Finally, we use our framework to reconstruct signaling pathways in breast cancer cells.

\section{Methods}
\subsection{Our concepts} An {\it Information Flow (IF)} is a directed linear path from one node to another node in signaling pathways which does not allow self transition or transition to a previously visited node. An {\it Information Flow Gene Set (IFGS)} is the set of all genes in an IF with a random permutation of their ordering. The length of an IFGS is the number of genes present in the set. Therefore, there are $L!$ putative information flows that are compatible with an IFGS of length $L$. We assume throughout that $L \geq 3$. An IF of length two serves as prior knowledge. Given a collection of $m$ unordered IFGS's $X_1,~X_2,\ldots,X_m$, we treat the order $\Theta_i$ associated with $X_i$ as a random variable. We write $(X_i,\Theta_i)$ to represent this association. Let us assume that the length of $X_i$ is $L_i$, for $i=1,\ldots,m$. As the sampling space of $\Theta_i$ corresponding to $X_i$ is of size $L_i!$, it follows that the sampling space of the joint distribution $P((X_1,\Theta_1),\ldots,(X_m,\Theta_m))$ is the set of $\prod_{i=1}^{m}L_i!$ permutations. Sampling space of size $\prod_{i=1}^{m}L_i!$ can be computationally intractable even for moderate values of $L_i$ and $m$. As a result, our goal of signaling pathway reconstruction can be translated into drawing sample signaling pathways sequentially from the joint distribution $P((X_1,\Theta_1),\ldots,(X_m,\Theta_m))$ (the true signaling pathway) of IFGS's and then estimating the most likely signaling pathway using sampled pathways.
\par Next, we present our two-stage GSGS framework. In Stage I, we derive IFGS's which form the building blocks of the signaling pathways. In Stage II,  we develop a Gibbs sampling like algorithm to sequentially sample permutation orders for each IFGS by conditioning on the remaining of the network structures.

\begin{figure}[!tpb]%figure1
\centering
\includegraphics[scale=0.31]{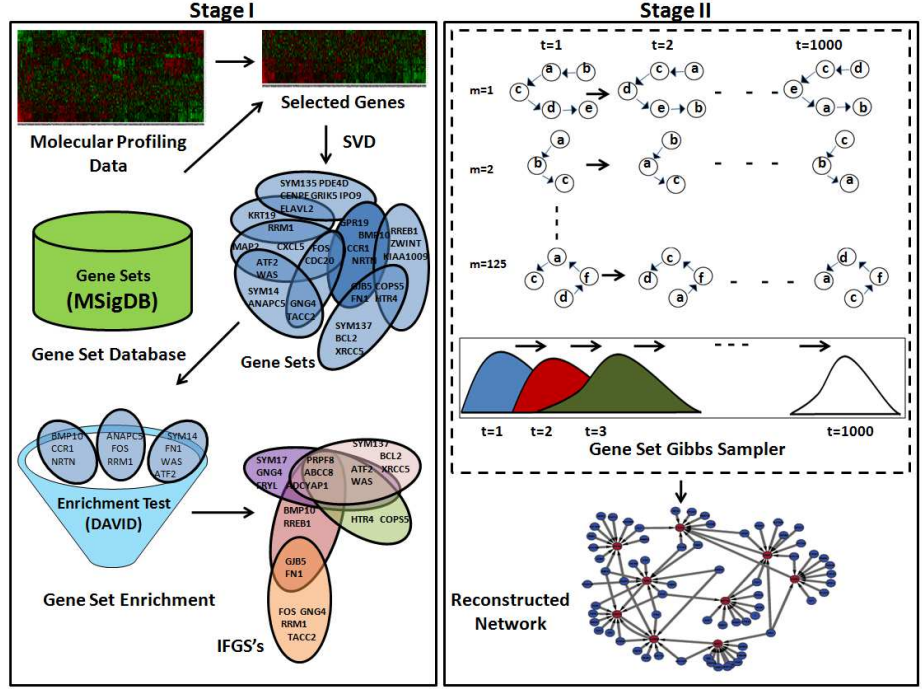}
\caption{\small{Flow chart for the two-stage GSGS signaling pathway reconstruction framework. Stage I: Derivation of IFGS's using two common data resources. Stage II: Gene Set Gibbs Sampler successively draws sample signaling pathways of the underlying true signaling pathway from the joint distribution of IFGS's.}}
\label{fig:f1}
\end{figure}

\subsection{Stage I: Derivation of IFGS's} In Stage I, we derive unordered and overlapping IFGS's corresponding to latent information flows to serve as input for the pathway reconstruction algorithm presented in the next section (Fig. \ref{fig:f1}). We use Singular Value Decomposition (SVD) to identify candidates gene sets. To extract coherent gene sets, the algorithm combines knowledge from two complementary forms of data, gene sets available from data bases and molecular profiling data from high-throughput platforms. We first select genes which appear most frequently in the gene set compendium under study. This frequency is referred to as \emph{degree}. We identify significant genes by fitting a power law distribution ($y\propto x^{-\alpha},~\alpha>1$) on the degrees of distinct genes present in the compendium. An application of SVD on the gene expression data $D$ corresponding to significant genes leads to a factorization of the form $D_{p\times q} = U_{p\times p} \cdot S_{p\times q} \cdot V_{q \times q}^T$, where $p$ is the number of genes and $q$ is the number of samples. We choose column vectors from $U$ corresponding to $k$ highest singular values in SVD. In general, $k$ is comparatively smaller than the original dimension of data. Following \citealt{Kim03}, we assume that $k$ satisfies $k(m+n)<mn$. We let $k = \max\{r: r(m+n)<mn\}$ to derive the maximum number of gene sets by preserving the preceding criteria. It is well known that a single gene in a living cell may simultaneously participate in multiple biological processes. The chosen basis vectors represent $k$ potential information flows. For a specified cut-off $\beta$, we set the top $\beta\%$ entries (in absolute values) among $k$ vectors as significant and other entries as zero. The non-zero locations in $k$ vectors correspond to $k$ overlapping gene sets. We further perform gene set enrichment analysis on the gene sets derived using SVD. The enriched gene sets represent IFGS's.
\subsection{Stage II: Signaling pathway reconstruction from IFGS's}
{\it Joint distribution and conditional distribution of gene sets}. With increasing number of gene sets, the size of the sampling space for the multivariate distribution $P((X_1,\Theta_1),\ldots,(X_m,\Theta_m))$ is of the order of $\prod_{i=1}^{m}L_i!$. Such a space might be computationally intractable even for moderate values of $L_i$ and $m$. However, it is possible to theoretically describe this distribution under certain assumptions.
\par Now onwards, we consider IFGS's as random samples from a first order Markov chain model, where the state of a node is only dependent on the state of its previous node. We compute the initial probability vector $\pi_0$ and transition probability matrix $\Pi$ from $m$ IF's (ordered paths) as follows. If there are a total of $n$ distinct genes across $m$ IF's, then
\begin{equation}
\pi_{0} = (\frac{c_1}{c},\ldots,\frac{c_n}{c})
\label{eq1}
\end{equation}
\noindent where $c_l$ is the total number of times $l^{th}$ gene appears as the first node among $m$ IF's, for each $l=1,\ldots,n$ and $c=\sum_{l=1}^n c_l$. If $c_{rs}$ is the total number of times $r^{th}$ gene transits to $s^{th}$ gene (i.e. there is edge from $r$ to $s$) among $m$ ordered paths, then
\begin{equation}
\Pi = [p_{rs}]_{n\times n}
\label{eq2}
\end{equation}
\noindent where $p_{rs} = c_{rs}/\sum_{s=1}^n c_{rs}$, $r,s=1,\ldots,n$.

\par The computation of $\pi_0$ and $\Pi$ allows us to calculate the likelihood of each of the $\prod_{i=1}^{m}L_i!$ collections of IF's. The likelihood of each collection is the product of the likelihoods of $m$ individual IF's in the collection. As each IF is treated as a first order Markov chain, we can calculate its likelihood using $\pi_0$ and $\Pi$. For example, we compute the likelihood of the information flow $z\rightarrow y \rightarrow x$
\begin{equation}
\mathcal P({z \rightarrow y \rightarrow x})= P(z)\times P(y|z)\times P(x|y).
\label{eq3}
\end{equation}
The likelihood values calculated for a total of $\prod_{i=1}^{m}L_i!$ collections of IF's can be normalized to denote a distribution of permutation ordering probabilities. However, the computation of $\prod_{i=1}^{m}L_i!$ likelihoods might be computational intractable. This serves as motivation for the proposed Gibbs sampling like approach. The computational tractability of our GSGS framework lies in sampling an order for each IFGS $X_i$ by conditioning on the orders of the remaining IFGS's, with a much reduced sample space of size $L_i!$.

\par Let us write all IFGS's and their associated orderings together as $(\overline X, \overline \Theta)$, where $\overline X = (X_1,\ldots,X_m)$ and $\overline \Theta = (\Theta_1,\ldots,\Theta_m)$. The notations are suffixed with $-i$ to consider all but the $i^{th}$ component, e.g. $\overline X_{-i}$, $(\overline X, \overline \Theta)_{-i}$ etc., for $i\in \{1,\ldots,m\}$. We sample an order for the $i^{th}$ gene set $X_i$ by conditioning on the known orders of remaining $m-1$ gene sets $X_1,\ldots,X_{i-1},X_{i+1},\ldots,X_m$. To sample an order for $X_i$ from the conditional distribution,
we leave the $i^{th}$ gene set out, and compute the initial probability vector $\pi_{-i}$ and transition probability matrix $\Pi_{-i}$ by following the procedure described in Eq. \ref{eq1} and Eq. \ref{eq2}, from $m-1$ IF's. Further, we calculate the likelihoods of all possible orders $\Theta_i^j,~j=1,\ldots,L_i!$ for $X_i$ by conditioning on the orders of remaining $m-1$ gene sets. The conditional likelihood for the $j^{th}$ order for $X_{i}$ is given by
\begin{equation}
\mathcal L_{i}^{j} = \begin{cases}
\frac{\mathcal P_{i}^j}{\sum_{j=1}^{L_i!}\mathcal P_{i}^j}
& \text{if $\sum_{j=1}^{L_i!}\mathcal P_{i}^j \neq 0$},\\
\frac{1}{L_i}& \text{otherwise}
\end{cases}
\label{eq4}
\end{equation}
where
\begin{equation}
\mathcal P_{i}^j = P((X_i,\Theta_i=\Theta_i^j)|(\overline {X},\overline{\Theta})_{-i}).
\label{eq5}
\end{equation}
\noindent For a fixed value of $j$, $\mathcal P_i^j$ is computed by decomposing it into the product of conditional probability terms. For example, we compute the likelihood of $z\rightarrow y \rightarrow x$ corresponding to the gene set $X_i = \{x,y,z\}$ as
\begin{equation}
\mathcal P((X_i,\Theta_i = {z \rightarrow y \rightarrow x})|(\overline {X},\overline{\Theta})_{-i}) = P(z)\times P(y|z)\times P(x|y).
\end{equation}
Each term on the right is conditioned on $(\overline {X},\overline{\Theta})_{-i}$ and is available from $\pi_{-i}$ and $\Pi_{-i}$. We now sample an order for $X_i$ from the conditional distribution using inverse Cumulative Density Function (CDF) (\citealt{Gelman03}). The CDF of the conditional distribution $P((X_i,\Theta_i)|(\overline {X},\overline{\Theta})_{-i})$ is defined as
\begin{equation}
F((X_i,\Theta_i=\Theta_i^j)|(\overline {X},\overline{\Theta})_{-i})) = \sum_{k=1}^{j} \mathcal P_i^k
\label{eq7}
\end{equation}

\begin{algorithm}
\caption{Gene Set Gibbs Sampler}
\begin{algorithmic}[1]
\label{algorithm1}
\STATE \textbf{Input:} $m$ IFGS's $X_i,~i=1,\ldots,m$, prior knowledge (optional), burn-in state $B$ and number of samples $N$ to be collected after burn-in state
\STATE \textbf{Output:} $m$ information flows $(X_i, \hat \Theta_i)$, $i=1,\ldots,m$
\STATE At $t=0$, make a random choice of order $\Theta_i^{(0)}$ from $L_i!$ permutations,~$i=1,\ldots,m$
\FOR {$t=1,\ldots,B+N$}
\STATE $\overline \Theta = (\Theta_1^{(t-1)},\ldots,\Theta_m^{(t-1)})^T$
\FOR {$i=1,\ldots,m$}
\STATE Compute $P^{(t)}_{-i}$ and $\Pi^{(t)}_{-i}$
\STATE Calculate the conditional likelihoods $\mathcal L^j_i$'s (Eq. \ref{eq4}) of $L_i!$ permutations by treating $X_i$ as a first order Markov chain
\STATE Draw an order $\Theta_i^{(t)}$ for $X_i$ from the conditional distribution $P((X_i,\Theta_i)|(\overline {X},\overline{\Theta})_{-i})$
\STATE Update the order information for $X_i$
\ENDFOR
%\STATE $\overline \Theta = (\Theta_1^{(t)},\ldots,\Theta_m^{(t)})^T$
\ENDFOR
\STATE Return $\hat \Theta_i = \text{mode}(\Theta_i^{(B+1)},\ldots,\Theta_i^{(B+N)})$,~$i=1,\ldots,m$.
\end{algorithmic}
\end{algorithm}
\noindent
for each $j=1,\ldots,L_i!$. By sampling a number $u\sim U(0,1)$ and letting $F^{-1}(u) = v$, we get a randomly drawn order $v$ for $X_i$ from the conditional distribution (Eq. \ref{eq7}).
\\\\\noindent
{\it Gene Set Gibbs Sampler}. In Algorithm \ref{algorithm1}, we present Gene Set Gibbs Sampler, which leads to the reconstruction of signaling pathways from IFGS's derived in Stage I. In case of prior knowledge, we augment known edges as directed pairs with unordered IFGS's, and keep the direction of these edges fixed during the execution of the algorithm. Algorithm \ref{algorithm1} outputs a list of IF's. To reconstruct signaling pathways, we start with an empty network of distinct genes present in the input list and reconstruct the most likely signaling pathway by joining IF's present in the output of Algorithm \ref{algorithm1}.

\subsection{Burn-in state} A burn-in state in Algorithm \ref{algorithm1} refers to a stage after which we start collecting sampled pathways. Samples collected after burn-in state are assumed to be drawn from the joint distribution of IFGS's. To determine an appropriate burn-in state, we translated the approach presented in (\citealt{Gelman03}, \citealt{Givens05}) in our framework to compute the ratio
\begin{equation}
R = \frac{\frac{N-1}{N} W_v + \frac{1}{N} B_v}{W_v}
\label{eq:r}
\end{equation}
\begin{algorithm}
\caption{Network2GeneSets}
\begin{algorithmic}[1]
\label{algorithm2}
\STATE \textbf{Input:} A directed acyclic graph with $n$ nodes
\STATE \textbf{Output:} All IFGS's
\FOR {$i=1,\ldots,n$}
\IF{node $i$ has no children}
\STATE continue
\ELSE
      \IF{node $i$ has children}
      \STATE add to Queue $Q$ and the Linked List $L$ all the directed pairs consisting of $i$ and a child of $i$
      \ENDIF
\WHILE {$Q$ is not empty}
\STATE Pop an information flow $P$ from $Q$
\IF {the last node in $P$, say $k$, has no children}
\STATE continue
\ENDIF
\STATE add to $Q$ and $L$, all information flows obtained by appending each child of $k$ to $P$
\ENDWHILE
\ENDIF
\ENDFOR
\STATE Prune information flows in $L$ of length 2 (prior knowledge)
\STATE Randomly permute orders of information flows in $L$ and order of genes in each information flow
\STATE Return all IFGS's of length $\geq 3$.
\end{algorithmic}
\end{algorithm}
\noindent for three quantities sensitivity, specificity and PPV. Here, $N$ is the total number of pathways sampled after burn-in state, $W_v$ is the averaged within-chain variance and $B_v$ is between-chain variance. Moreover, Sensitivity = TP/(TP+FN), Specificity = TN/(TN+FP) and PPV = TP/(TP+FP), where TP = number of true positives, TN = number true negatives, FP = number of false positives, and FN = number of false negatives. In practice if $\sqrt R < 1.2$, the choice of burn-in state and $N$ is acceptable (also see \emph{Supplementary Material}).
\subsection{Computational complexity} The worst case time complexity of Gene Set Gibbs Sampler is $Nm(m+n+FL)$, where $N$ is the number of sampled pathways, $m$ is the number of IFGS's, $n$ is the number of distinct genes, $L$ is the length of the longest gene set in the input and $F = L!$. As longer gene sets $(L\geq 10)$ are less likely to correspond to information flows, the complexity arising from $FL$ could be managed by appropriately selecting the length of gene sets in Stage I. Thus, the computational complexity of our algorithm increases quadratically with increase in the number of IFGS's, which compares very favorably with the Bayesian network approaches.
\section{Data Analysis}
We analyzed the performance of our proposed network inference framework by reconstructing three different gene regulatory networks. We obtained two gold standard directed networks from the {\it In Silico} Network Challenge in DREAM initiative.
The two networks are {\it In Silico} network (\citealt{Mendes09,Stolovitzky09}) from DREAM2 and {\it E. coli} network (\citealt{Marback09}, \citealt{Marback10}, \citealt{Prill10}) from DREAM3 network challenges. {\it E. coli} and {\it In Silico} networks consist of $50$ nodes, with $62$ and $37$ true edges respectively. Availability of gold standard networks allows us to assess the performance of the proposed approach. In addition, we also implemented our two-stage GSGS framework to reconstruct signaling pathways in breast cancer cells.
\subsection{Derivation of IFGS's}
From the \emph{E. coli} and \emph{In Silico} networks, two collections of IFGS's were derived by a direct application of Algorithm \ref{algorithm2}. Indeed, Algorithm \ref{algorithm2} finds all unordered gene sets from a given network. The algorithm first finds all IF's (linear paths) in the network and then randomly permutes the ordering of genes in each IF. We may note that Algorithm \ref{algorithm2} is more general than the standard Depth First Search (DFS) algorithm in that the latter does not find all the linear paths. There were a total of $125$ and $57$ IFGS's of length $\geq 3$ for the \emph{E. coli} and \emph{In Silico} networks, respectively. These collections of IFGS's serve as input for Gene Set Gibbs Sampler (Algorithm \ref{algorithm1}).
\par  We also derive IFGS's using the C4 gene set compendium (computational gene sets) from MSigDB (\citealt{Subra05}). There are a total of $883$ overlapping cancer gene sets and $10,124$ distinct genes in the compendium. We identified significant genes $(P(X\geq x) \geq 0.95)$ by fitting a power law distribution on the degrees of $10,124$ genes (Fig. 6, {\it Supplementary Material}). We obtained a total of $289$ genes using this selection procedure. We also collected $299$ samples of breast cancer patients from Affymetrix HG-U133 plus 2.0 platform. A total of $267$ out of $289$ selected genes could be mapped to the annotation table for the Affymetrix HG-U133 plus 2.0 platform. For each of the $267$ genes, gene expression levels corresponding to exactly one probe set with highest average measurement among $299$ samples were selected. The resulting data set contained $267$ rows (genes) and $299$ columns (samples). We performed SVD on the breast cancer gene expression data of size $267\times 299 ~(m\times n)$ and considered $k$ basis vectors corresponding to $k$ highest eigenvalues. As mentioned in Section 2, we chose $k$ by setting $k = \max\{r: r(m+n)<mn\} = 141$. To identify the most significant candidates for IFGS's, top $2\%$ of the  entries across $k$ basis vectors were declared as non-zero and the remaining entries were set as zero. We derived a total of $138$ candidate gene sets by identifying genes corresponding to non-zero entries among $k$ basis vectors. We lost 3 gene sets by constraining a gene set to contain at least 3 genes. To measure the enrichment of gene sets, we further performed gene set enrichment analysis using the functional annotation tool in DAVID (\citealt{Dennis03}, \citealt{Huang09}). DAVID performs gene set enrichment analysis using a modified Fisher Exact Test. We used Affymetrix Human Genome U133 Plus 2.0 Array as background to test the enrichment of gene sets. By setting the other parameters in DAVID as default, $106$ enriched gene sets containing a total of $212$ distinct genes were derived. The enriched gene sets serve as IFGS's.
\begin{figure}[!tpb]
\centering
\includegraphics[scale=0.31]{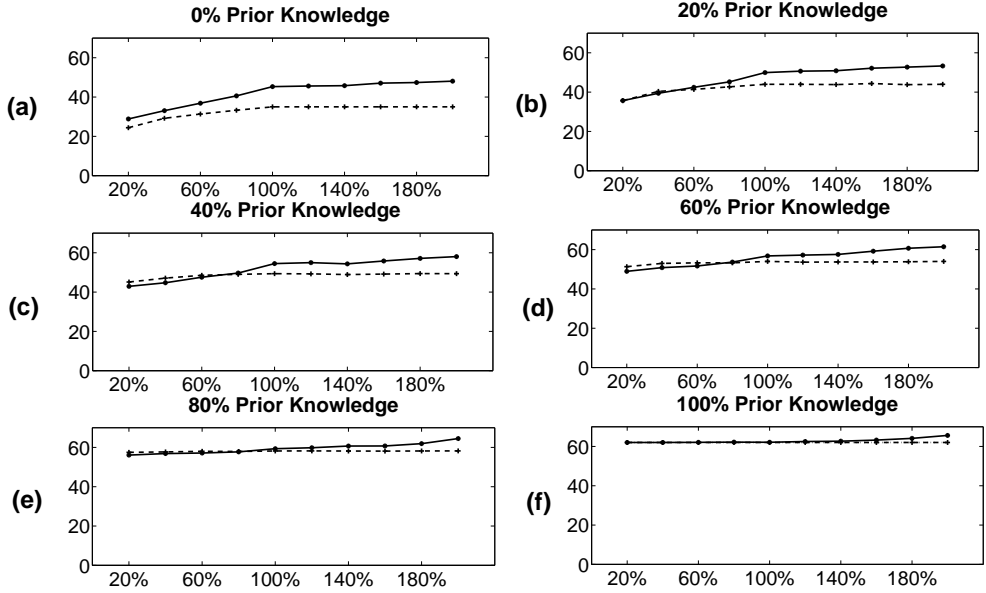}
\caption{\small{Sensitivity analysis for the GSGS approach with increasing percentage of prior knowledge. Network: \emph{E. coli}. In blocks (a)-(f), $x$-axis represents the percentage of gene sets present in the input and $y$-axis plots the total number of edges predicted by GSGS (solid line).  The dashed line plots correspond to the ground truth. Here, we have considered only those genes which were present among IFGS's after pruning all gene pairs.}}
\label{fig: f3}
\end{figure}
\subsection{Performance evaluation using \emph{E. coli} network}
We now analyze the performance of Gene Set Gibbs Sampler using \emph{E. coli} network. Analogous results for {\it In Silico} network are presented as {\it Supplementary Material}. Using Gene Set Gibbs Sampler (Algorithm \ref{algorithm1}), we collected a total of $500$ networks after burn-in state which we fixed at $500$. As all gene pairs are pruned by Algorithm \ref{algorithm2}, some genes might be lost and never appear in the input list of IFGS's. We compare the network predicted by Algorithm \ref{algorithm1} with the subnetwork formed by genes present in the input. A detailed list of settings is presented in the \emph{Supplementary Material}. With the chosen set of parameters, $\sqrt R$ in Eq. \ref{eq:r} was found approximately equal to one, for each of the three quantities sensitivity, specificity and PPV. We used the total number of predicted true edges and F-score to assess the performance of Algorithm \ref{algorithm1}. The F-score is defined as $F=2pr/(p+r)$. Here, $r$ is the sensitivity and $p$ is the PPV.
\vskip 0.20cm
\begin{table}
\begin{center}
\begin{tabular}{|c|c|c|c|c|c|c|}
\hline
   & 0\% & 20\% & 40\% & 60\% & 80\% & 100\% \\
   \hline
  20\% & 0.430 & 0.648 & 0.748 & 0.844 & 0.926 &1\\
40\% & 0.496	& 0.680 & 0.792	& 0.865 & 0.937 & 1\\
60\% & 0.513 & 0.677 &	0.790 &	0.883 &	0.943 &	1\\
80\% & 0.468 &	0.665 &	0.780 &	0.860 &	0.947 &	0.999 \\
100\% & 0.457 &	0.595 &	0.719 &	0.824 &	0.923 &	0.999 \\
120\% & 0.459 & 0.590 &	0.704 & 0.825 &	0.913 &	0.996\\
140\% & 0.450 &	0.579 &	0.722 &	0.805	& 0.909 & 0.999 \\
160\% & 0.422 &	0.564 &	0.691 &	0.803 &	0.913 &	0.991 \\
180\% & 0.434 &	0.550 &	0.679 &	0.786 &	0.897 &	0.984 \\
200\% & 0.425 &	0.546 &	0.676 &	0.778 &	0.877 & 0.974\\
\hline
\end{tabular}
\end{center}
\vskip 0.20cm
\caption{\small{F-scores calculated for the GSGS approach with increasing percentage of gene sets in the input (row) and prior knowledge (column). Network: \emph{E. coli}. We observe a clear increasing trend in the F-scores in each row, indicating the positive impact of incorporating prior knowledge, while a clear trend of similarity is observed within each column, indicating a marked robustness of the performance of GSGS to the over-sampling and under-sampling of gene sets.}}
\label{table1}
\end{table}
\par In order to accommodate the real-world under-sampling and over-sampling situations, we first performed sensitivity analysis of the GSGS approach using \emph{E. coli} network. Fig. \ref{fig: f3} demonstrates the effect of removing and adding unordered gene sets to the input list of IFGS's in Algorithm \ref{algorithm1}. In Fig. \ref{fig: f3}, $x$-axis represents the percentage of gene sets present in the input list, where $20\%$ means that $80\%$ of the gene sets were randomly removed from the list of all IFGS's, and $120\%$ means that $20\%$ of randomly sampled gene sets were added to the original list of all IFGS's. In Fig. \ref{fig: f3}, we present the performance of our approach in terms of the total number of predicted true edges. In blocks (a)-(f), the number of edges identified by the GSGS approach remains close to the ground truth. We also observe the positive effect of incorporating prior knowledge. As the percentage of prior knowledge increases (block (a) to block (f)), difference between the ground truth and prediction decreases. In particular, our approach does not produce a large number of false positives in the presence of redundant gene sets.
\par In Table \ref{table1}, we present the F-scores for the GSGS approach with increasing percentage of gene sets (rows) and prior knowledge (columns). We observe that the F-scores increase with an increase in the percentage of prior knowledge (values in a row), and these scores remain close on removal or addition of gene sets (values in a column) demonstrating an impressive robustness to under-sampling and over-sampling. This observation strongly supports the applicability of our GSGS framework in the real-world scenarios, where we often do not observe all gene sets or the observed gene sets are redundant.
\begin{figure}[!tpb]
\centering
\includegraphics[scale=0.30]{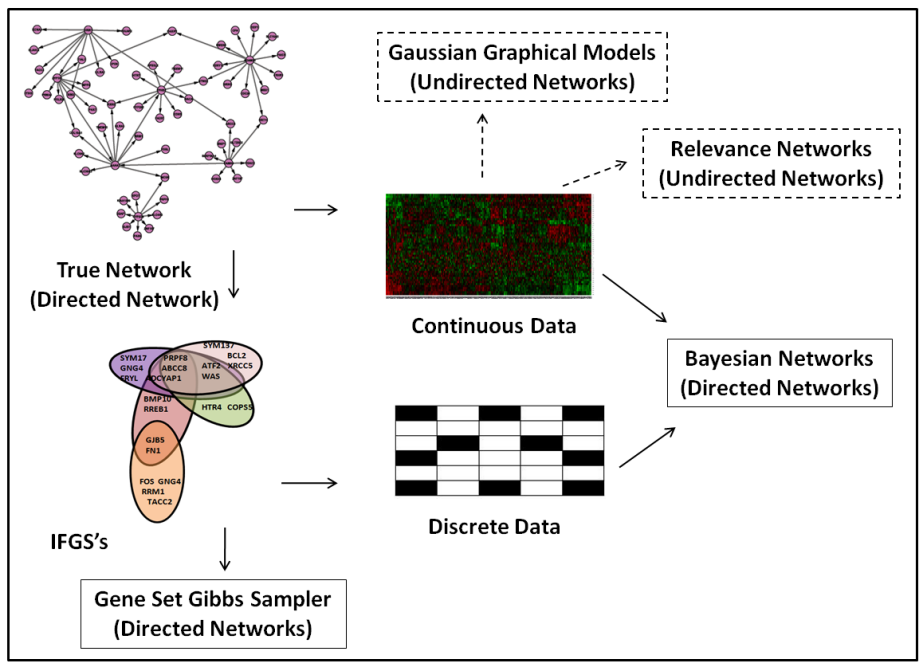}
\caption{\small{A sketch of the idea behind comparing the GSGS approach with Bayesian network approaches. Note that the underlying network from which gene sets are derived is a directed network. Moreover, gene sets can equivalently be represented as a matrix of binary discrete values.
Bayesian networks are the best choice in this case to fairly assess the performance of GSGS. Bayesian network approaches accommodate both discrete and continuous data, and reconstruct a directed network.}}
\label{fig:f2}
\end{figure}
\par We also compare the performance of our approach with a number of popular network inference approaches (\citealt{Margolin06}, \citealt{Meyer08}) with a primary emphasis on the two Bayesian network approaches, K2 and MCMC (Metropolis-Hastings or MH) implemented in the Bayes Net Tool Box (BNT) (\citealt{Murphy01b}b, \url{http://sourceforge.net/projects/bnt/files/}). The main reasons are the following: (1). From methodology point of view our method infers the most probable linear structure(s) using likelihood scores calculated from the products of conditional probabilities. It is essentially in the same sprit as Bayesian network approaches while fundamentally different from other approaches based on the calculation of pair-wise similarity. (2). Both our approach and Bayesian network approaches naturally take discrete data in that a collection of gene sets can equivalently be represented as a matrix of binary discrete values. Indeed, each IFGS naturally corresponds to a binary sample derived by considering the presence and absence of a gene in the set. Most of the existing network reconstruction algorithms are more suitable for inferring an undirected network from continuous data sets.
\par In Fig. \ref{fig:f2}, we sketch the idea behind comparing our approach with the Bayesian network approaches. Our goal in this paper is to infer the underlying directed network. Also note that a collection of gene sets can be represented as a matrix of binary discrete values. A binary sample corresponding to an IFGS can be derived by assigning a value $0$ to the genes not present in the IFGS and $1$ otherwise. Bayesian network approaches can accommodate both discrete and continuous data sets and reconstruct a directed network. The equivalent representation of gene sets as binary discrete data makes the comparison between our gene set based approach and the Bayesian network approaches very fair. In addition, we also generated continuous data to serve as input for the Bayesian network and other approaches (\citealt{Margolin06}, \citealt{Meyer08}). Thus, using the {\bf same underlying network}, e.g. the \emph{E. coli} network, as the sole input (Fig. \ref{fig:f2}): (1). We generate discrete data inputs for Gene Set Gibbs Sampler (Algorithm \ref{algorithm1}) by collecting IFGS's in the output of Algorithm.
(2). We generate discrete data inputs for K2 and MH by considering the absence ($0$) or presence ($1$) of a gene in each IFGS in the output of Algorithm \ref{algorithm2}. (3). We generate continuous data inputs for K2, MH and MINET using BNT.
\begin{figure}[!tpb]
\centerline{
\includegraphics[scale=0.26]{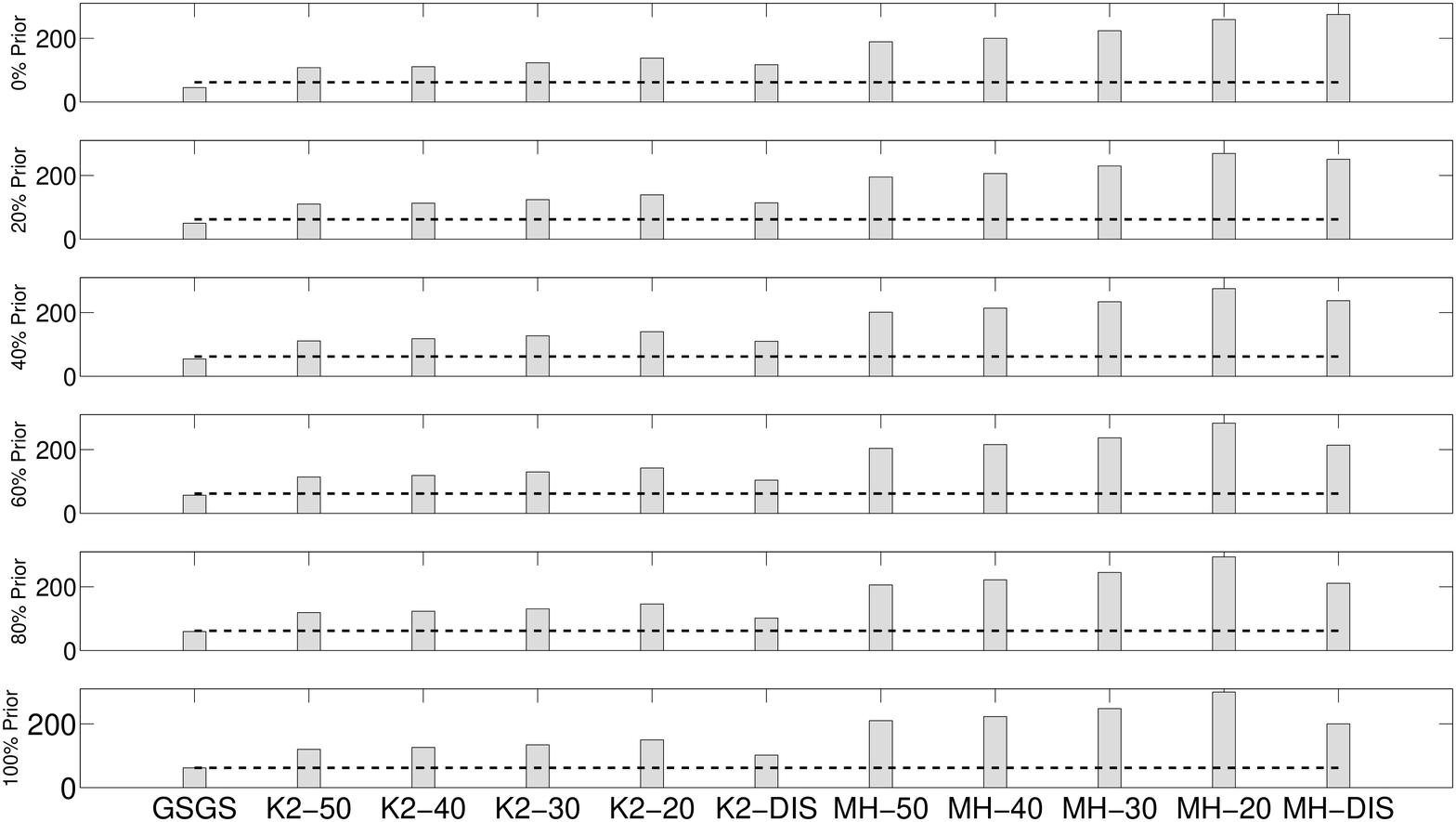}
}
\centerline{
\includegraphics[scale=0.26]{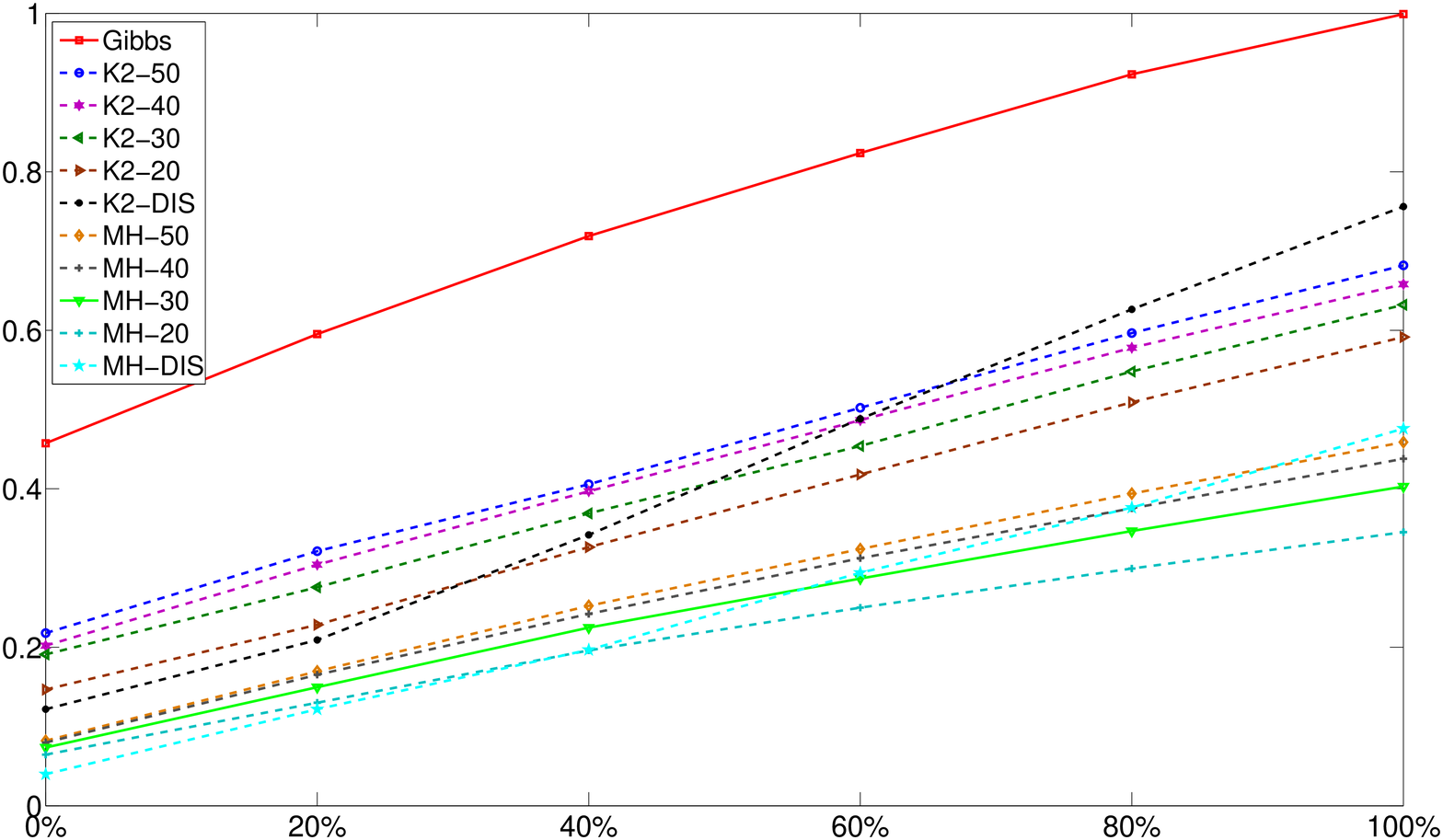}
}
\caption{\small{Network: \emph{E. coli}. (Upper Panel) Comparison of the GSGS  approach with K2 and MH in terms of Total Number of Predicted Edges with increasing percentage of prior knowledge. In each panel ``Method-N" stands for a Bayesian network method applied to continuous data of sample size N, and ``Method-DIS" corresponds to using binary discrete data. Bayesian Information Criterion (BIC) and Bayesian scoring were used on the corresponding data sets. The dashed line represents ground truth. (Lower Panel) Comparison of the GSGS  approach with K2 and MH in terms of F-score. Here $x$-axis represents the percentage of prior knowledge and $y$-axis plots F-scores from three approaches.}}
\label{fig: f4}
\end{figure}
\begin{figure*}[!tpb]
\centering
\includegraphics[scale=0.25]{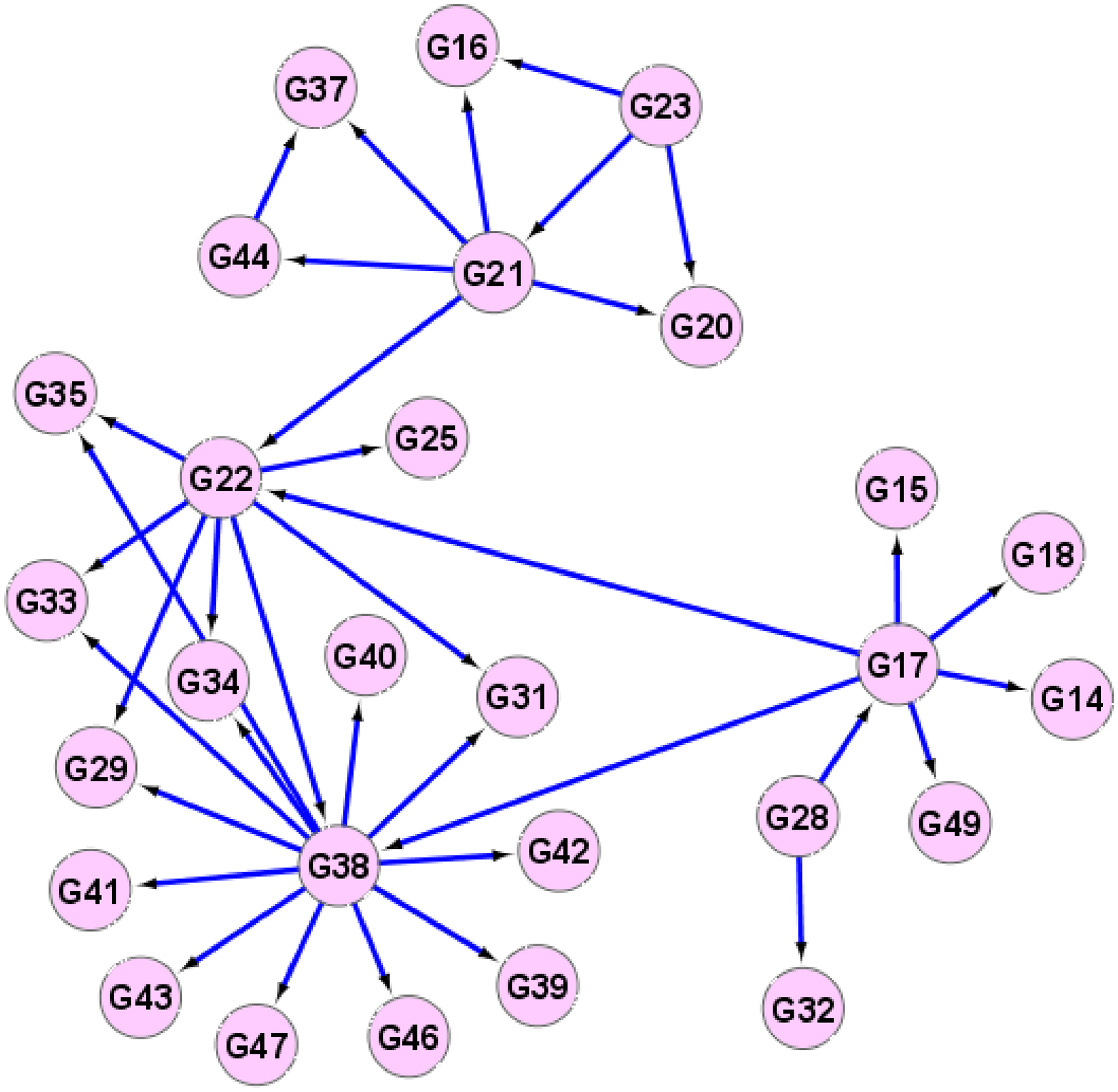}
\includegraphics[scale=0.25]{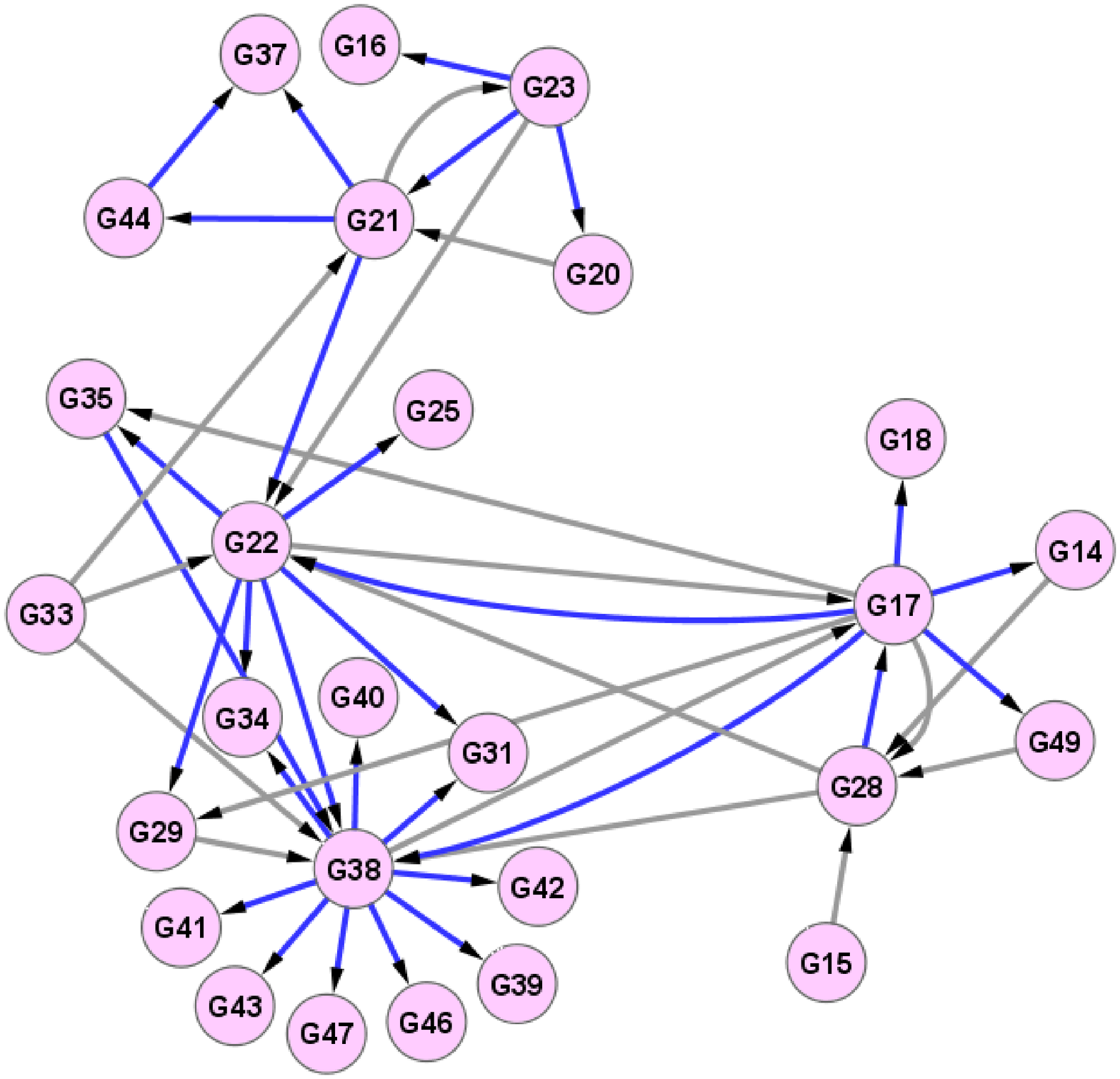}
\includegraphics[scale=0.26]{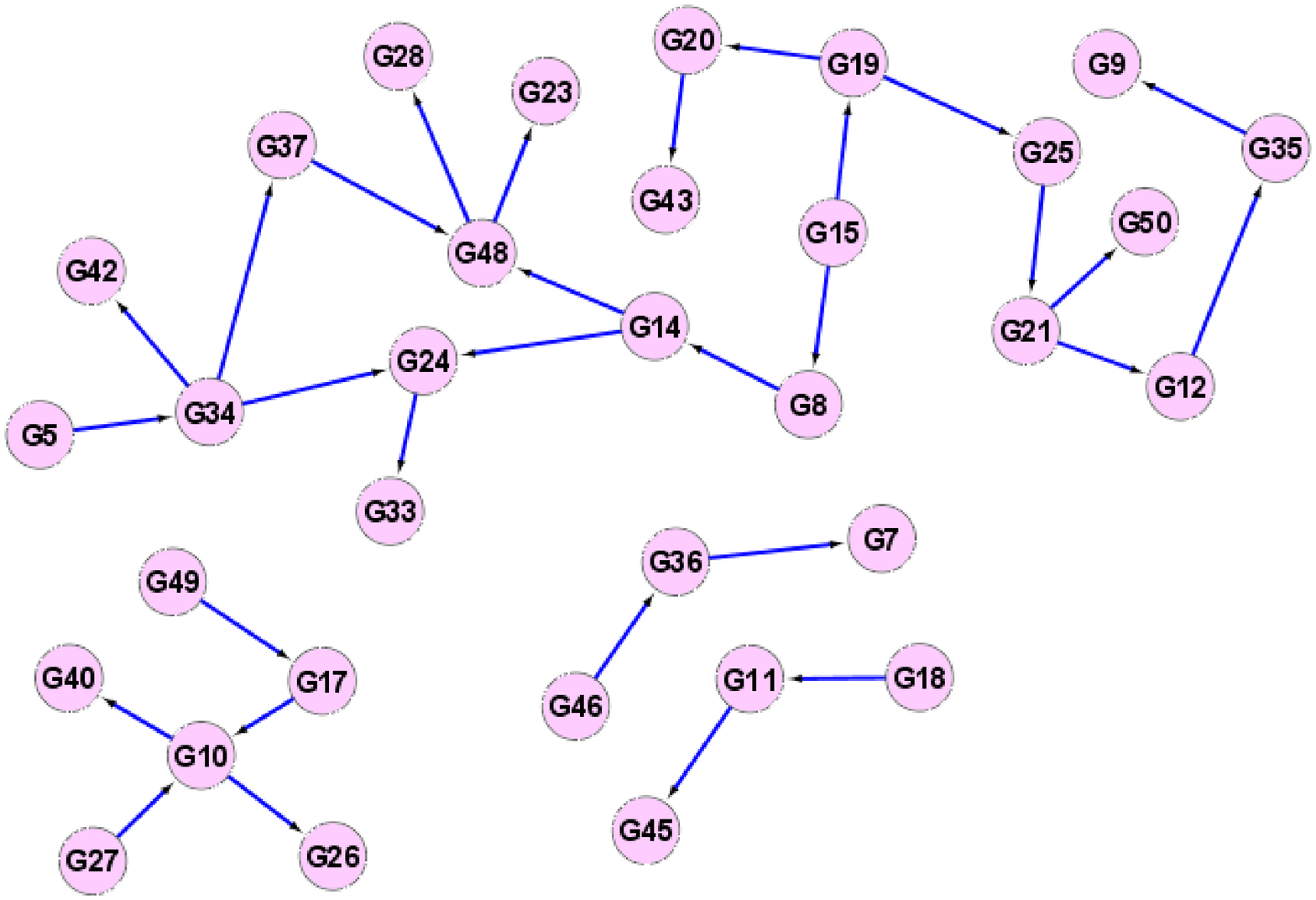}
\includegraphics[scale=0.26]{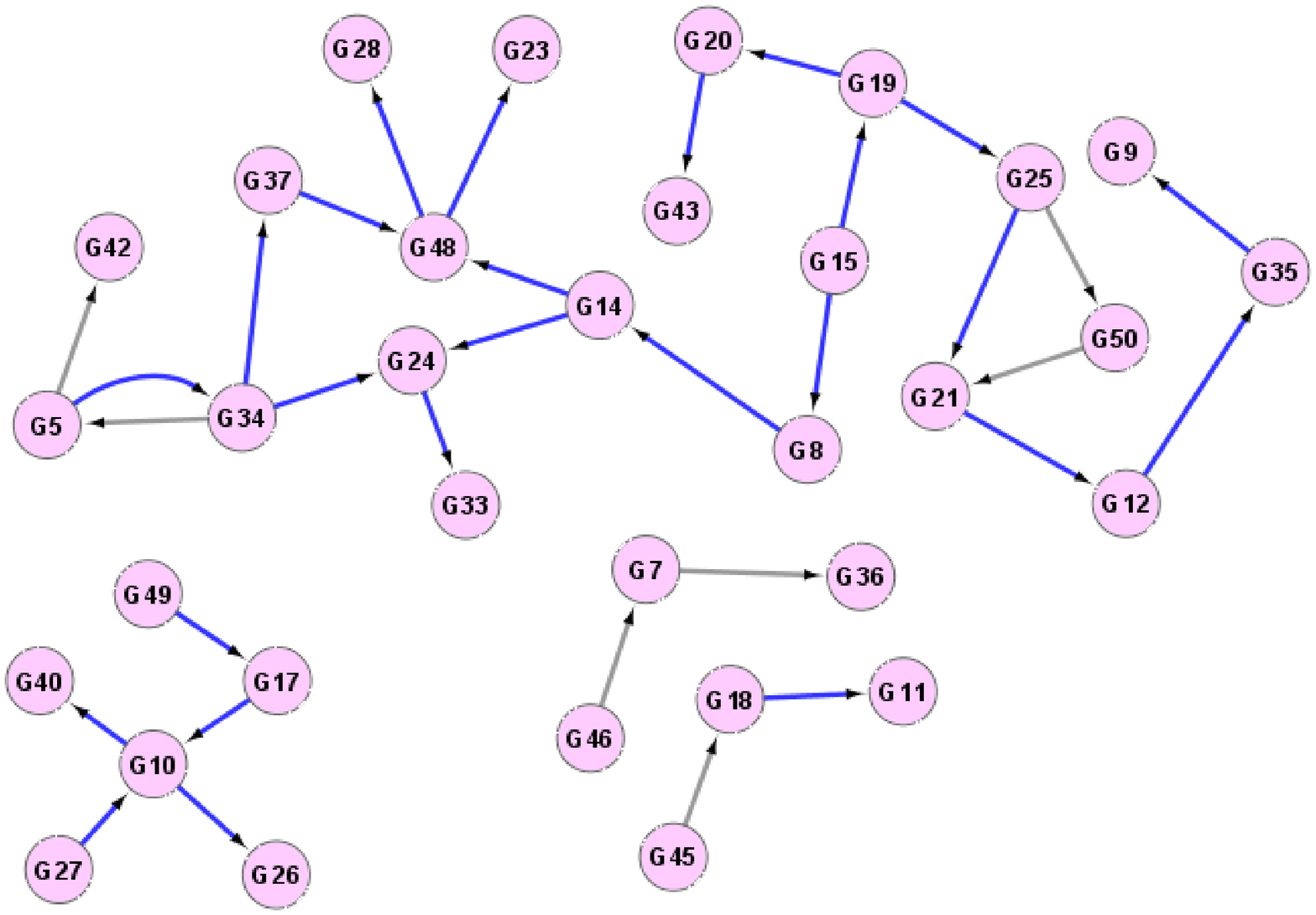}
\caption{\small{A proof of principle study. Left panels show two gold standard networks, {\it E. coli} (Upper) and {\it In Silico} (Lower); Right panels show the corresponding predicted networks by GSGS, {\it E. coli} (Upper) and {\it In Silico} (Lower). For a fair comparison, all stand-alone linear paths of length $2$ are removed from both networks. On the right panels, the blue edges correspond to true positives and gray edges represent false positives. Figures were generated using Cytoscape (\citealt{Shannon03}).}}
\label{fig: f5}
\end{figure*}
\par In principle, the K2 approach (\citealt{Cooper92}) first specifies an ordering of nodes involved in the underlying network. Thus, initially each node has no parent. The algorithm incrementally assigns a parent to a node whose addition increases the score of the resulting structure the most. For the $i^{th}$ node, parents are chosen from the set of nodes with index $1,\ldots,i-1$. On the other hand, the MH algorithm (\citealt{Murphy01a}a) starts from an initial directed acyclic network $G_0$ and selects a network $G_1$ uniformly from the neighborhood of $G_0$. The neighborhood of a network $G$ is the collection of all directed acyclic networks which differ from $G$ by addition, deletion or reversal of a single edge. The algorithm accepts or rejects the move from $G_0$ to $G_1$ by computing an acceptance ratio defined in terms of marginal likelihood ratio $P(D|G_1)/P(D|G_0)$. Here $D$ represents the given data. This procedure is iterated starting from the most recent network. A specified number of networks are collected after burn-in state. For scoring a structure, BNT implements Bayesian Information Criterion (\citealt{Schwartz78}) and Bayesian score functions (\citealt{Cooper92}).
\begin{table}
\begin{center}
\begin{tabular}{|c|c|c|c|c|c|}
  \hline
   & GSGS & CLR & ARACNE & MRNET & MRNETB \\\hline
  {\it E.coli} & 0.457 & 0.230 & 0.377 & 0.303 & 0.228 \\\hline
  {\it In Silico} & 0.431 & 0.238 & 0.425 & 0.389 & 0.327 \\
  \hline
\end{tabular}
\end{center}
\vskip 0.20cm
\caption{\small{Performance comparison of GSGS with four other pair-wise similarity based network reconstruction approaches using F-scores. The sample size is $50$. }}
\label{table2}
\end{table}
\par In the upper panel of Fig. \ref{fig: f4}, we plot the results from a comparative study in terms of total number of predicted edges. It is clear that K2 and MH predict many false positives. In the lower panel of Fig. \ref{fig: f4}, we have plotted the F-scores for different approaches with increasing percentage of prior knowledge. We observe that F-scores for the GSGS approach is significantly higher than K2 and MH. Further, the impact of incorporating prior knowledge on F-score is more prominent in case of GSGS than K2 and MH. F-scores for both K2 and MH remain much lower than the GSGS approach even in the presence of a large amount of prior knowledge. For similar results using \emph{In Silico} network, we refer to the \emph{Supplementary Material}. We also compare GSGS with four other approaches without using prior knowledge. The F-score results are presented in Table \ref{table2}.
In Figure \ref{fig: f5}, we provide more detailed evidence of the superior performance of our method using both \emph{In Silico} and \emph{E coli} networks. In Figure \ref{fig: f5}, two left panels represent the true topologies of both networks, and two right panels represent the reconstructed network topologies using GSGS. In each reconstructed network, blue edges represent true positives and gray edges represent false positives. A high level of accuracy is observed in both the reconstructed networks.
\subsection{Pathway Reconstruction in Breast Cancer Cells}
Before using the IFGS's for signaling pathway reconstruction, we validated our underlying assumption that a large network is built from unordered and overlapping IFGS's. We measured the amount of overlapping among IFGS's. Indeed, we computed the number of genes shared by different number of gene sets (Fig. 7, {\it Supplementary Material}). A minimum of $75\%$ of total genes were found to be shared by at least two IFGS's. An exponentially truncated power law distribution ($y\propto x^{-\alpha}e^{-\beta x}$) was fitted on the degrees of genes (Fig. 8, {\it Supplementary Material}). Such networks naturally occur in biology (\citealt{Ghaz06}).
\begin{figure}[!tpb]%figure1
\centering
\includegraphics[scale=0.50]{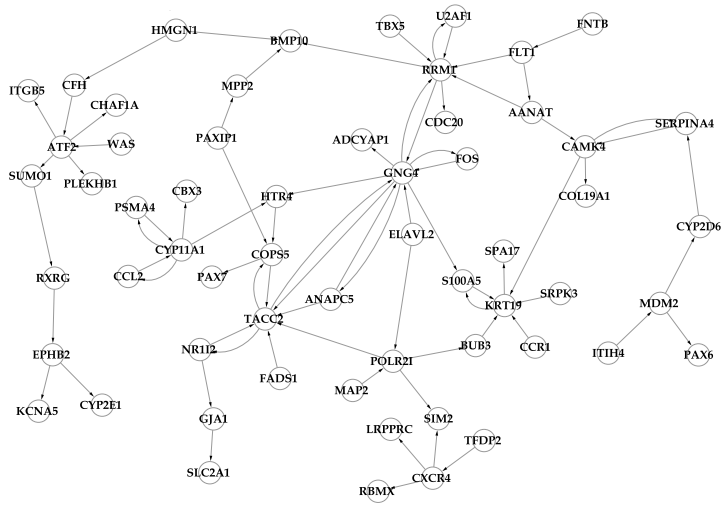}
\caption{\small{A partial view of the subnetwork formed by nodes with a minimum of five first order neighbors in the network reconstructed from the genes related to breast cancer. Figure was generated using Cytoscape (\citealt{Shannon03}).}}
\label{fig: f6}
\end{figure}
\par A total of $20$ candidate signaling pathways from $20$ independent runs of Algorithm \ref{algorithm1} were predicted. To summarize a single network, we declared all the edges appearing in at least $5$ networks as true edges, for a fair compromise between sensitivity and PPV (Fig 9, {\it Supplementary Material}). In Fig. \ref{fig: f6}, we present a subnetwork formed by nodes with at least 5 first order neighbors in the reconstructed network. Indeed, nodes with high connectivity are likely to participate in many signaling transduction events. We made use of GeneCards (\citealt{Safran10}) to verify the relevance of genes in the subnetwork with breast cancer and signaling events. We found that many genes, e.g. BMP10, CCL2, CCR1, COL19A1, CXCR4, EPHB2, FLT1, FOS, GNG4, ITGB5 and MDM2 shown in Fig. \ref{fig: f6}, are involved in the molecular mechanisms of cancer. In addition, MDM2 is involved in HER-2 signaling in breast cancer, POLR2I in hereditary signaling in breast cancer and ATF2 in Estrogen-dependent breast cancer signaling(Sigma-Aldrich \url{www.sigmaaldrich.com}). CXCR4 is highly expressed in breast cancer cells (\citealt{Muller01}, RefSeq \url{www.ncbi.nlm.nih.gov/refseq}) whereas GJA1 is marker for detecting early oncogenesis in the breast (Genatlas \url{http://genatlas.medecine.univ-paris5.fr}). RRM1 is located in the imprinted gene domain of 11p15.5 (an important tumor-suppressor gene region). Alterations in this region are associated with breast cancer (RefSeq). ATF2 and its two direct neighbors WAS and ITGB5 participate in CDC42 pathway (Applied Biosystems Pathway \url{www.appliedbiosystems.com}). Similarly, BMP10 and HMGN1 are involved in ERK signaling, and EPHB2 and KCNA5 in PI3K signaling. Genes appearing in the directed path from FLT1 to EPHB2 via BMP10 and ATF2, and genes in the path from GNG4 to EPHB2 via  BMP10 and ATF2, are highly relevant to MAPK signaling and P38 signaling. For example, BMP10 is connected to ATF2 by a linear path. It has been reported that TAK1 and the SMAD pathways activated by BMPs activate several transcription factors like ATF2 (\citealt{Monzen01}). Similarly, FLT1 and GNG4 which are closely situated and connected by a linear path, have been reported to participate in many signaling events, e.g. ERK signaling, PI3K Signaling, P38 signaling and MAPK signaling. These evidences further support the use of GSGS framework for signaling pathway reconstruction.
\section{Conclusion}
In this paper, we proposed a novel computational framework, GSGS, to reconstruct signaling pathways from gene sets. As far as we know, the proposed framework is original in the following aspects: (1). It offers a unique two-stage framework for network reconstruction by combining knowledge from existing gene sets and molecular profiling data from high-throughput platforms (2). The ordering of genes in each gene set is treated as a random variable to capture the higher order interactions among genes participating in signal transduction events. In most of the existing approaches, individual genes are treated as variables (3). The problem of signaling pathway reconstruction is cast into the framework of parameter estimation for a multivariate distribution. (4). The true signaling pathways are modeled as a probability distribution of sample signaling pathways.
\par We first assessed the performance of our network inference algorithm by using two gold standard networks: {\it E.coli} and {\it In Silico}. Our approach was shown to have significantly better performance in terms of F-score and total number of predicted edges than the Bayesian network and other pairwise similarity based approaches (\citealt{Margolin06}, \citealt{Meyer08}). Robustness of our approach against under-sampling or over-sampling of gene sets was proved by performing sensitivity analysis. We applied our GSGS framework to reconstruct a network in breast cancer cells, and verified it using existing database knowledge. Overall, our analyses favor the use of our two-stage GSGS framework in the inference of complicated signaling pathways.
\par The advent of systems biology has been accompanied by the blooming of network construction algorithms, many of which treat gene pairs as the basic building block of the signaling pathways and reconstruct signaling pathways by simultaneously detecting co-expressed gene pairs using molecular profiling data (e.g. \citealt{Butte03}, \citealt{Zhu05}, \citealt{Margolin06}, \citealt{Meyer08}). This type of approaches enjoy simplicity and a much alleviated computational load but gene pairs do not represent the entire signal transduction pathways. Other approaches heuristically search for the higher scored network structure(s), such as Bayesian networks (e.g. \citealt{Cooper92}, \citealt{Song09}). Many network structures may be found to be statistically plausible, but similar to the gene pairs they do not necessarily represent the real signaling transduction mechanisms. Moreover, the computation loads of searching for a higher scored network is prohibitively high and a number of assumptions on the network structures have to be made, such as small size of the parent sets. Our GSGS framework infers the most likely signaling pathway(s) from a probability distribution of sampled signaling pathways using overlapping gene sets inferred from molecular profiling data. The reconstructed information flows are faithful representation of the real-world signaling transduction mechanisms. The advantages of gene set based computational approaches have been adequately demonstrated in the many bioinformatics research areas, for example, disease classification and enrichment analysis, we expect our gene set based GSGS framework to open a new avenue in methodology research of signal transduction.

\section*{Acknowledgments}
This work was supported by NIH grant R21LM010137 to D.Z.

\end{document}